\documentclass[prd,aps,showpacs,two column]{revtex4}
\usepackage{latexsym}
\usepackage{amsmath}
\usepackage{amssymb}
\usepackage[all]{xy}
\usepackage[dvips]{graphicx}

\newcommand{\be}{\begin{equation}}
\newcommand{\ee}{\end{equation}}
\newcommand{\bn}{\begin{eqnarray}}
\newcommand{\en}{\end{eqnarray}}
\newcommand{\p}{\partial}
\newcommand{\nn}{\nonumber\\}

\newcommand{\dslash}{\partial \hspace{-1.7mm} /}

\begin{document}

\title{Monopoles in the presence of the Chern-Simons term via the Julia-Toulouse Approach}
\author{L. S. Grigorio}
\email{leogrigorio@if.ufrj.br} \affiliation{Instituto de F\'\i sica, Universidade
Federal do Rio de Janeiro, 21945, Rio de Janeiro, Brazil}
\author{M. S. Guimaraes}
\email{marceloguima@gmail.com} \affiliation{Instituto de F\'\i sica, Universidade
Federal do Rio de Janeiro, 21945, Rio de Janeiro, Brazil}
\author{C. Wotzasek}
\email{clovis@if.ufrj.br} \affiliation{Instituto de F\'\i sica, Universidade Federal
do Rio de Janeiro, 21945, Rio de Janeiro, Brazil}


\begin{abstract}

We study $QED_3$ with magnetic-like defects using the Julia-Toulouse condensation mechanism (JTM) introduced in \cite{Quevedo:1996uu}. By a careful treatment of the symmetries we suggest a geometrical interpretation for distinct debatable issues in the MCS-monopole system: (i) the induction of the non-conserved electric current together with the Chern-Simons term (CS), (ii) the deconfinement transition and, (iii) the computation of the fermionic determinant in the presence of Dirac string singularities. The JTM leads to proper interpretation of the non-conserved current as originating from Dirac brane symmetry breaking. The mechanism behind this symmetry breaking is clarified. The physical origin of the deconfinement transition becomes evident in the low energy effective theory induced by the JTM. The proper procedure to compute the fermionic determinant in the presence of Dirac branes will be presented. A byproduct of this approach is the possible appearance of statistical transmutation and the clarification for the different quantization rules for the topological mass.

\end{abstract}

\maketitle


\section{Introduction and motivations}

It seems to be a well established fact that the presence of a Chern-Simons term destroys current conservation in the 3D Maxwell theory with a magnetic charge (an instanton in this dimensionality). This non-conserved electric current is located at the Dirac brane (a generalization of the Dirac string \cite{Dirac} appearing in $4D$; in this $3D$ case it is a point) associated with the monopole. Conservation can be regained by the {\it ad hoc} introduction of an extra electric current term so that the total electric current is conserved \cite{Henneaux:1986tt}.

Although it seems clear that, for some reason, the Dirac brane attached to the magnetic instanton acquires electric charge and becomes {\it visible} in the spectrum of the theory \cite{Pisarski:1986gr}, the physical origin of such phenomenon -- magnetic instanton emitting electric charge -- has remained obscure for many years (for some further discussion see \cite{Lee:1991ge}). To our understanding, the induction of the Chern-Simons (CS) term from the fermionic condensate \cite{Redlich} and the electric current induced from the magnetic instanton {\it are not} distinct phenomena and to treat them as so is the sole cause of the interpretation difficulties faced in previous studies.

It is the main purpose of this study to break with this lore by showing that both the CS term and the electric charge of the instanton Dirac brane are induced from the quantum fluctuations of the fermionic fields in such a manner that the overall current conservation is maintained. This claim comes from interpreting the quantum fluctuations as a condensate \cite{Redlich, Gamboa:2008ne}. This will allow us to employ a very useful prescription to establish the phenomenology of the resulting system. It is known as the Julia-Toulouse mechanism \cite{JT} and it was generalized in \cite{Quevedo:1996uu} by Quevedo and Trugenberger for general $p$-form fields, where they have shown that such approach is able to describe the Higgs mechanism from a dual perspective as a defect condensation phenomenon. In fact a pioneering take on these matters was done in \cite{Kleinert:1982dz} (see also \cite{Kleinert:1989kx}).

Instrumental in this construction is a proper treatment of the full set of local symmetries. It has been shown in \cite{Kleinert} that in the presence of both electric and magnetic charges, besides the usual (electric) gauge symmetry, there appears another set of gauge-like symmetries associated with the presence of the magnetic branes. They will be called collectively brane symmetry and will play a decisive role in the results that follow.

We will clearly show that the introduction of highly massive fermionic matter interacting with the Maxwell field in the presence of magnetic defects is responsible for the breaking of the brane symmetry along with the induction of the CS term. However, a new and unexpected term will also come along which will be responsible for a possible new physics. This will be described in a dual formulation where the brane symmetry breaking is viewed as a consequence of the condensation process.  A proper treatment of the brane symmetry in this process will result in a suitably redefinition of the original $QED_3$ into the Maxwell-Chern-Simons (MCS) theory in the presence of defects. This constitutes the continuum formulation of the lattice system studied in \cite{Diamantini:1993iu} and clarifies the deconfinement phenomenon that occurs in the 3D Maxwell-monopole system upon the introduction of the CS term. Furthermore the formalism seems to suggest the new possibility of anyonic statistics for the total conserved electric current besides clarifying why there are many different (Dirac) quantization rules for the topological mass. The whole procedure gives the proper way to formulate and compute fermionic determinants in the presence of Dirac branes, an issue still subject to debate.

In the next section we will discuss some known results reviewing the brane symmetry concept and the interpretation of the quantum fermionic fluctuations as a condensation effect. Next these concepts will be applied to the problem of defining the MCS theory in the presence of magnetic defects. This will lead us to the Henneaux-Teitelboim (HT) \cite{Henneaux:1986tt} formulation plus an extra spin term furnishing statistical qualifications to the conserved total electric current, a feature not present in the original HT formulation.

\section{Branes and condensation}

\subsection{Brane symmetry}

Before establishing our main results, it will pay off to review and clarify some fundamental concepts. In this section we will discuss how a proper treatment of the brane symmetry associated with external sources can lead to important conclusions.

Consider Maxwell theory in $(2+1)D$ in the presence of external magnetic poles. These are instantons in this dimensionality having a pseudo-scalar current $\rho$ defined by:
\begin{eqnarray}
\label{m01}
\partial_{\mu} {}^* F^{\mu} =g \rho,
\end{eqnarray}
where ${}^* F^{\mu}$ is the Hodge dual of the electromagnetic field strength and $g$ is the magnetic coupling. Eq.(\ref{m01}) stands for a violation of the Bianchi identity preventing us to naively define a vector potential. This can be accomplished however if we write the magnetic current as
\begin{eqnarray}
\label{m02}
\rho = \partial_{\mu} {}^* \Omega^{\mu}.
\end{eqnarray}
The current $\rho$ is defined as a classical external source, so ${}^* \Omega^{\mu}$ is a Dirac-like magnetic brane. Then eq. (\ref{m01}) can be written as
\begin{eqnarray}
\label{m03}
\partial_{\mu} \left({}^* F^{\mu} - g {}^* \Omega^{\mu}\right) = 0,
\end{eqnarray}
which is naturally solved introducing a vector potential leading to the electromagnetic field strength in the presence of magnetic sources given by
\begin{eqnarray}
\label{m04}
F_{\mu\nu} = \partial_{\mu} A_{\nu} - \partial_{\nu} A_{\mu} + g  \Omega_{\mu\nu}.
\end{eqnarray}
For $F_{\mu\nu}$ to be a physical observable it must be invariant under any ambiguity introduced in the formulation. The familiar gauge invariance is built in
\begin{eqnarray}
\label{m05}
A_{\mu} \rightarrow A_{\mu} - \partial_{\mu} \chi ; \;\;\; F_{\mu\nu} \rightarrow  F_{\mu\nu}.
\end{eqnarray}
There is however another important gauge ambiguity that was introduced when the magnetic brane was defined \cite{Kleinert}. Looking at (\ref{m02}) it is clear that the magnetic current (an observable) is invariant under
\begin{eqnarray}
\label{m06}
{}^* \Omega^{\mu} \rightarrow {}^* \Omega^{\mu} + \varepsilon^{\mu\nu\rho}\partial_{\nu}L_{\rho}
\end{eqnarray}
For $F_{\mu\nu}$ to be invariant too, $A_{\mu}$ must respond to (\ref{m06}) as
\begin{eqnarray}
\label{m07}
A^{\mu} \rightarrow A^{\mu} - g L^{\mu} \, .
\end{eqnarray}
In a lattice formulation one usually starts with the free Maxwell theory and demands this property for the gauge field $A_{\mu}$, which is then properly recognized as a compact variable (an angular variable), as a result the magnetic poles appear naturally as defects of the gauge field. Here we are considering these defects as external inputs. Consistency in the formulation demands this new transformation property for the gauge field. There is, as we shall see, much to be gained by considering this transformation as independent of the usual gauge transformations. We will call it brane transformation. This corresponds of course to the Dirac string construction \cite{Dirac} properly generalized for the present dimensionality. We are following here the same formulation discussed in \cite{Kleinert}.

\begin{figure}[hbt]
       \centering  
       \includegraphics[scale=0.3]{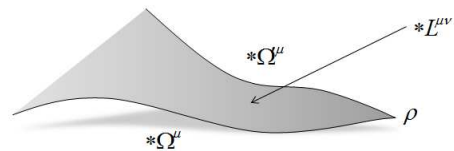}
       \caption{\it lines ${}^* \Omega'^{\mu}$ and ${}^* \Omega^{\mu}$ bordering the surface ${}^* L^{\mu\nu}$. These lines meet at the instanton location $\rho$.}
       \label{fig:1}
   \end{figure}

If we further add an electric current $J^{\mu}$, the system becomes defined by the Bianchi identity (\ref{m01}) and the equation of motion
\begin{eqnarray}
\label{m08}
\partial_{\mu} F^{\mu\nu} = eJ^{\nu}= e\epsilon^{\nu\rho\alpha}\partial_\rho \Lambda_\alpha
\end{eqnarray}
that follows from the action
\be
\label{p10}
S = \int d^3x\left[- {\frac 14} \left(\partial_{[\mu} A_{\nu]} + g \Omega_{\mu\nu}\right)^2 - e A_\mu J^\mu(\Lambda_\alpha)\right]
\ee
where the branes defined by $\Lambda_\alpha$ and $\Omega_{\mu\nu}$ play the role of (extended) electric and magnetic charges, respectively.
The invariance of the system under (\ref{m06}) and (\ref{m07}) can be checked using the Dirac quantization condition $eg=2\pi n; {\rm for} \,\, n= {\rm integer}$. Another way to see that, which is more convenient for our purposes, is to cast the action in terms of brane invariants only. This is in consonance with a new approach introduced in \cite{Chernodub:2008rz} to consider spontaneous symmetry breaking. The brane transformation (\ref{m06}) may be interpreted geometrically as follows: the line defined by ${}^* \Omega^{\mu}$ is deformed into another configuration ${}^* \Omega'^{\mu}$. These lines border a surface which is defined by ${}^* L^{\mu\nu}$ (see fig.\ref{fig:1}).

This suggests the following definition. Let ${}^*\tilde \Omega^{\mu}$ be a fixed brane defined by
\begin{eqnarray}
\label{m09}
{}^*\tilde \Omega^{\mu} = {}^* \Omega^{\mu} - \varepsilon^{\mu\nu\rho}\partial_{\nu}K_{\rho}.
\end{eqnarray}
where, under the transformation (\ref{m06}), $K_{\mu}$ responds as
\begin{eqnarray}
\label{m10}
K_{\mu} \rightarrow  K_{\mu} + L_{\mu},
\end{eqnarray}
so that ${}^*\tilde \Omega^{\mu}$ is invariant. This is just a change of variables. All the information contained in ${}^* \Omega^{\mu}$ is now in the combination ${}^* \tilde \Omega^{\mu} + \varepsilon^{\mu\nu\rho}\partial_{\nu}K_{\rho}$ with all the magnetic gauge redundancy being now carried by $K_{\mu}$. In terms of these new variables the action (\ref{p10}) becomes
\begin{eqnarray}
\label{m11}
S &=& \int d^3x- {\frac 14} \left(\partial_{[\mu} \tilde A_{\nu]} + g \tilde \Omega_{\mu\nu}\right)^2 + e \tilde A_\mu J^\mu(\Lambda_\alpha)\nonumber\\
 &-& eg K_\alpha \epsilon^{\alpha\mu\nu}\p_\mu \Lambda_\nu
\end{eqnarray}
where $\tilde A^{\mu} =  A^{\mu} + g K^{\mu}$ is the brane invariant vector potential. Observe that the last term in (\ref{m11}) is the only one which is not brane invariant because of (\ref{m10}). This term represents a brane-brane interaction between the electric and magnetic sources. It is an intersection number and therefore an integer requiring the Dirac charge quantization to avoid its contribution quantum mechanically.

\subsection{Chern-Simons term as a condensation effect}

It is well known that the Chern-Simons term may be considered as an effect of the quantum fluctuations of massive fermionic matter. It was pointed out by Redlich \cite{Redlich} (some of these ideas were already present in \cite{Niemi:1983rq}) that if the fermions are very massive they are dynamically inert but their quantum fluctuations may drastically alter the vacuum of the electromagnetic field. This occurs because in $(2+1)D$ massive fermions in its minimal representation necessarily breaks P and T symmetries. The picture painted by Redlich to explain this symmetry breaking is that of a condensate of polarized spins filling up the vacuum which ends up inducing in the electromagnetic field an angular momentum current which is just the Chern-Simons term. In fact, in $(2+1)D$ the fermionic mass term is just a spin density operator and its vacuum expectation value is the Chern-Simons term
\begin{eqnarray}
\label{m12}
M <\bar \psi \psi> = M<\psi^{\dagger} \sigma_z \psi> = \frac{e^2}{4\pi} \frac{M}{|M|} {}^* F^{\mu}A_{\mu}
\end{eqnarray}

In \cite{Gamboa:2008ne} we have used this picture to construct a chain of connections between the free Maxwell theory, the massless scalar theory, the MCS theory and the SD theory. It was argued there that the radiative corrections known to generate the Chern-Simons term have a better physical interpretation in the dual picture as a condensation phenomenon that takes place generating the CS term. In the dual picture the condensing matter appears as defect singularities of the massless scalar field (which is dual to the Maxwell field in $(2+1)D$). After condensation the effective theory describing the excitations of the condensate is the SD theory which is the dual of the MCS theory \cite{Deser:1984kw}.
Observe however that in the original radiative corrections picture there is no phase transition in the usual way. The very presence of the fermionic field defines the ``new phase''. Another way of saying that is to claim that the phase transition in the original picture occurs for arbitrarily small couplings.

The prescription used to derive this effective theory is the Julia-Toulouse mechanism (JTM) \cite{JT} in its relativistic form described by Quevedo and Trugenberger \cite{Quevedo:1996uu}. The JTM amounts to the construction of an effective field theory in which the (magnetic) defects of the original field eventually condense. The details about the phenomenon that drives the condensation are not addressed nor they need to be because with very
general assumptions an unique form for the effective theory after the condensation may be constructed. Quevedo and Trugenberger considered as the only assumptions that the resulting effective theory was built up as a derivative expansion in the new field with respect to the new scale defined by the characteristic density of the condensate, possessing relativistic invariance, respecting the symmetries of the system and, most important, the need to recover the original model in the appropriate dilute limit.

The result obtained in \cite{Gamboa:2008ne} goes as follows. We start with the scalar field theory in the presence of defects
\begin{eqnarray}
\label{m13}   {\cal L}_{\phi} = \frac 12 \left( \partial_{\mu} \phi - e\Lambda_{\mu}\right)^2.
\end{eqnarray}
Brane symmetry is realized here by the transformations
\begin{eqnarray}
\label{m14}   \Lambda_{\mu} &\rightarrow& \Lambda_{\mu} + \partial_{\mu} \theta \nonumber\\
               \phi &\rightarrow& \phi + e\theta
\end{eqnarray}
So an obvious brane invariant is
\begin{eqnarray}
\label{m15}  f_{\mu}=\partial_{\mu} \phi - e\Lambda_{\mu}.
\end{eqnarray}
We are searching for a theory describing the physics of a $P$ and $T$ symmetry breaking condensate. The condensation process is a proliferation of the defects. A defect means a singularity in the scalar field, that is, the scalar field is not well defined at
the position of a defect and its Dirac brane. As the condensation process becomes energetically favored the scalar field becomes more and more singular until it is not defined anywhere and only the brane invariant field $f_{\mu}$ (\ref{m15}) retains any physical meaning. It describes the excitation field of the condensate. The Julia-Toulouse prescription prompt us to add terms to the lagrangian to account for the dynamics of
these excitations. The first such term in a derivative expansion that breaks the $P$ and $T$ symmetries is the CS term. So we arrive at the following effective description of the system after condensation of defects takes place
\begin{eqnarray}
\label{SD} {\cal L}_{SD} = \frac 12 f_{\mu}f^{\mu} +
\frac{1}{2m}f_{\mu}\varepsilon^{\mu\nu\rho}\partial_{\nu} f_{\rho}
\end{eqnarray}
which is just the SD theory, where $m$ is interpreted as the density on the condensate. Here we see that $m$ is proportional to $e^2$ by dimensional analysis. It is important to observe a characteristic general signature of the JT prescription: the ``rank-jump'' effect; that is, in the present case the scalar field has turned into a vector field.
Another important point to observe is that we can recover the free scalar field theory in the limit $m \rightarrow 0$. In this limit the last term in (\ref{SD}) forces the identification $\varepsilon^{\mu\nu\rho}\partial_{\nu} f_{\rho} = 0$ leading to $f_{\mu}=\partial_{\mu}\phi$ which turns (\ref{SD}) into the free scalar Lagrangian.

\section{Brane invariance and induced electric current}

In this section we want to use the ingredients gathered in the previous sections to construct the MCS theory in the presence of monopoles. The rationale goes as this: we know that the MCS is dual to the SD theory and by the general properties of duality we know that a minimal coupling is dualized into a non-minimal coupling. So if we start with the minimally coupled SD theory we will be able, by duality, to reach the sought for formulation of the non-minimally coupled MCS theory. But this is not as straightforward as it seems. The problem is that if we naively start with a minimally coupled SD theory we will not have brane symmetry from the beginning (this will be evident from the discussion below) and we would not get any information about how the brane symmetry was actually broken. To deal with this problem we will take another route. We will start with the scalar theory in the presence of defects that will eventually condense but we will also add a minimal coupling to the scalar theory. This will lead us after condensation to a well defined minimally coupled SD theory (meaning that it is defined only in terms of brane invariants) which after duality will give the properly defined MCS theory in the presence of magnetic defects.

One would think that the MCS plus magnetic defects system could be reached starting with the Maxwell theory in the presence of such defects and adding massive fermionic degrees of freedom which, by quantum fluctuations, would generate the searched for theory. The problem with this procedure is that the fermionic determinant is not well defined in the presence of Dirac branes (the calculations done in \cite{Fradkin:1990xy} was performed supposing a 't Hooft-Polyakov like defect, that is, without Dirac strings). Therefore, a proper computation of the fermionic determinant in the presence of Dirac monopoles remains an open problem. Incidentally a possible answer for this problem will be suggested from our approach.

The Julia-Toulouse condensation process, in turn, have an interesting physical consequence giving a new and proper interpretation for a well known result; it makes clear why in the presence of a CS term both the condensation of magnetic monopoles and the confinement of electric charges, are suppressed \cite{Affleck:1989qf}. It is just because the monopoles themselves are confined since they are inside an electric condensate already. This simple physical scenario is one of the greatest advantages of the dual view through the Julia-Toulouse mechanism.

To formally establish this result we will start with the massless scalar field theory in the presence of defects and minimally coupled to an external source
\be
\label{p60}
{\cal L}_{\phi} = \frac 12 \left(\p_\mu\phi - e \Lambda_\mu\right)^2 - g \phi \rho \, .
\ee
From here we cannot proceed just yet with the JT prescription since the last term does not form a brane invariant (with respect to $\Lambda_\mu$). To remedy this we write as before $\rho = \partial_{\mu} {}^* \Omega^{\mu}$, integrate by parts the last term and introduce the new brane $K_\mu$ (as discussed in (\ref{m09}), (\ref{m10}) and (\ref{m11}))
\bn
\label{p90}
\phi \rho &=& \p_\mu\phi {}^*\Omega^\mu \to \left(\p_\mu\phi - e \Lambda_\mu\right)
\left({}^*\Omega^\mu - \epsilon^{\mu\nu\rho}\p_\nu K\right)\nn
&=& f_\mu {}^*\tilde \Omega^\mu
\en
where the brane $K_\mu$, transforms as in (\ref{m10}). Observe that the added terms do not have any physical consequences; one of them contributes a total derivative and the others only contribute with an integer number times $2\pi$ if we consider that the quantization condition (\ref{m08}) is valid as it is. Now we may proceed with the JT prescription as explained in the last section obtaining
\be
\label{p110}
{\cal L}_f = \frac 12 f_\mu^2 + \frac 1{2m} f_\mu \epsilon^{\mu\nu\rho}\p_\nu f_\rho + gf_\mu {}^*\tilde \Omega^\mu \, .
\ee
As expected, the scalar field $\phi$ has {\it jumped the rank} into a vector field $f_\mu$, satisfying the self-dual dynamics plus a minimal coupling to the {\it non-conserved} current ${}^*\tilde \Omega^\mu$ defined in (\ref{m09}). In fact this is now a physical brane. It is, by construction, invariant under brane transformations and furthermore it carries energy. That this is so may be seen by eliminating the SD field $f_\mu$ from the action (\ref{p110}) which leads to an effective action as,
\be
\label{p125}
S_{eff} = \int d^3x \frac{g^2}{2}{}^*\tilde\Omega_\mu \left[\frac{m\epsilon^{\mu\alpha\nu}\p_\alpha + \partial^{\mu}\partial^{\nu} - g^{\mu\nu} m^2}{\p^2 + m^2}\right] {}^*\tilde\Omega_\nu
\ee
This expression clearly shows that the Dirac-brane has acquired physical reality since now it cost energy to realize brane transformations. In a sense the brane has become {\it thick}. If we remember its definition (\ref{m02}) (observe that $\rho=\partial_{\mu} {}^* \tilde \Omega^{\mu} = \partial_{\mu} {}^* \Omega^{\mu}$) we understand clearly what it represents; it is just the flux tube connecting the monopoles (by \emph{Poincare duality} $\rho$ is the border of ${}^* \tilde \Omega^{\mu}$) and expression (\ref{p125}) just means that there is a string tension and a confining potential given by the third term as discussed, for instance,  in \cite{Diamantini:1993iu}. As so it must now display a {\it minimal length} between the positions of the monopoles so as to minimize energy. Notice that the dilute phase limit ($m \to 0$) naturally displays the Coulomb potential for the monopole charges $\rho$, as it should.

The action (\ref{p110}) is the dual representation of the theory studied by HT. In order to verify this last statement we introduce the {\it master} Lagrangian
\be
\label{p130}
{\cal L}_f \to {\cal L}_{f,B} = \frac 12 f_\mu^2 +  B_\mu \epsilon^{\mu\nu\rho}\p_\nu f_\rho - \frac{m}2 B_\mu \epsilon^{\mu\nu\rho}\p_\nu B_\rho + g f_\mu {}^*\tilde \Omega^\mu
\ee
which can be seen to reduce back to (\ref{p110}) upon integration of the auxiliary field $B_\mu$. On the other hand, integrating out the original field $f_\mu$ gives us its dual representation as
\be
\label{p140}
{\cal L}_B =- \frac 12 \left(\epsilon^{\mu\nu\rho}\p_\nu B_\rho + g {}^*\tilde\Omega^\mu\right)^2 - \frac m2 B_\mu \epsilon^{\mu\nu\rho}\p_\nu B_\rho
\ee
Observe that this action is brane invariant. In fact, the brane symmetry is hidden. This is in tune with Elitzur's theorem \cite{Elitzur:1975im} and the general principles of symmetry breaking. Remember that in the usual $U(1)$ Higgs mechanism a gauge field is combined with a scalar field forming a new gauge invariant massive vector field which is the relevant degree of freedom in a particular energy scale (for a recent discussion along these lines see \cite{Chernodub:2008rz}). This is known as spontaneous symmetry breaking, a misleading name perhaps since the gauge symmetry is not actually broken but hidden. In the present situation the same kind of phenomenon is happening except that it is the brane symmetry that is spontaneously broken.

\begin{figure}[hbt]
       \centering  
       \includegraphics[scale=0.3]{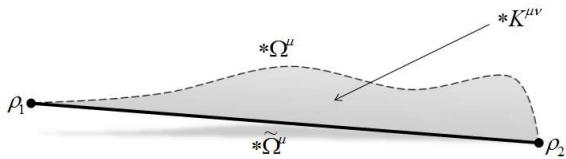}
       \caption{\it Two instantons $\rho_1$ and $\rho_2$ connected by a physical confining brane ${}^*\tilde \Omega^{\mu}$. Overall current conservation is accomplished taking into account the induced non-physical (brane non-invariant) electric current ${}^* \Omega^{\mu}$}
       \label{fig:2}
   \end{figure}

We can now make contact with the HT-theory \cite{Henneaux:1986tt} which is expressed in terms of a brane non-invariant gauge field. Observe that $B_\mu$ is a gauge field but it is brane invariant. In order to recover the description in terms of a brane non-invariant gauge field $A_\mu$ we observe that under transformations (\ref{m10}) and (\ref{m07}) for $K_\mu$ and $A_\mu$, respectively, the redefinition
\be
\label{p150}
B_\mu = A_\mu + g K_\mu
\ee
keeps the gauge field $B_\mu$ brane invariant. Bringing this redefinition back into (\ref{p140}) gives,
\begin{eqnarray}
\label{p160}
{\cal L}_A &=& \frac 12 \left(\epsilon^{\mu\nu\rho}\p_\nu A_\rho +g {}^*\Omega^\mu\right)^2 + \frac m2 A_\mu\epsilon^{\mu\nu\rho}\p_\nu A_\rho \nonumber\\
  &+& m g A_\mu\epsilon^{\mu\nu\rho}\p_\nu K_\rho + \frac{mg^2}{2} K_\mu\epsilon^{\mu\nu\rho}\p_\nu K_\rho.
\end{eqnarray}
The first three terms comprise the HT action \cite{Henneaux:1986tt} while the last term is new. This last term is a spin term and will be discussed momentarily. Observe first that the total current minimally coupled with the gauge field is conserved as it should. From eq.(\ref{m09}) we see that it may be expressed as the difference of two terms: ${}^*\tilde \Omega^{\mu}$ is just the non-conserved electric current introduced in \cite{Henneaux:1986tt} to compensate for the non-conservation of the induced electric current ${}^* \Omega^{\mu}$. Here they have a very clear geometrical interpretation, they are the border of the brane $K^{\mu}$ (see fig.\ref{fig:2}).

As for the spin term we can write it as
\begin{eqnarray}
\label{m16}
\rightarrow\int d^3x a_\mu\epsilon^{\mu\nu\rho}\p_\nu K_\rho  - \frac{1}{2mg^2}\int d^3x a_\mu\epsilon^{\mu\nu\rho}\p_\nu a_\rho
\end{eqnarray}
where $a_\mu$ is a statistical ``fictitious'' field \cite{Wilczek:1990ik}. This tells us that the total conserved electric current obeys a statistics given by the parameter
\begin{eqnarray}
\label{m17}
 \theta = \frac{mg^2}{2}.
\end{eqnarray}
If $\theta = 0 \;mod\;2\pi$ the excitations are bosons. If $\theta = \pi \;mod\;2\pi$ they are fermions. For any other value they are anyons.

In \cite{Henneaux:1986tt} a quantization of the topological mass was obtained. It was found that
\begin{eqnarray}
\label{m18}
 mg^2 = 2\pi n; \;\;\; n\in Z,
\end{eqnarray}
therefore the current can be bosonic or fermionic depending on $n$. This is a new non-trivial conclusion forced on us by the brane symmetry considerations just discussed. If the current has fermionic character this term will contribute to the partition function with a modulating phase. On the other hand Pisarski in \cite{Pisarski:1986gr} obtained a more stringent condition
\begin{eqnarray}
\label{m19}
 \frac{mg^2}{2} = 2\pi n; \;\;\; n\in Z.
\end{eqnarray}
and therefore in this case the current is bosonic for all $n$ and the last term does not contribute.

In our formalism mass quantization seems not to be demanded though. The result obtained in \cite{Henneaux:1986tt} was based on the observation that the total conserved electric charge coupled with the gauge field $A^{\mu}$ has charge $mg$ as can be read in (\ref{p160}) and thus, the argument goes, as any other electric charge it must be submitted to the Dirac quantization condition which leads to (\ref{m18}). Here however we have seen that the Dirac brane becomes observable so the usual Aharonov-Bohm like argument leading to charge quantization does not seem to apply after condensation has taken place. The result obtained in \cite{Pisarski:1986gr} is based on the compactification of the time direction (that is finite temperature) and arguing with the non trivial gauge transformations in this scenario, so it is also a different set than the one considered here. Furthermore in our approach the mass parameter is a phenomenological entity characterizing the resulting condensate so there is no a priori restriction in its value. As a result it may be possible for the excitations to be anyons.

As a final comment observe that our result suggests that the fermionic coupling with the Maxwell field in the presence of defects should be defined with respect to the brane invariant field $B^{\mu}$ figuring in (\ref{p140}), not the gauge field $A_\mu$. That is, the action (\ref{p140}) can be obtained through fermionic radiative corrections starting with the Maxwell theory in the presence of defects, with $B^{\mu}$ as gauge field, coupled with a massive fermionic field as given by the Dirac action as
\begin{eqnarray}
\label{m20}
\int d^3 x \bar{\psi}(i\dslash - e {B \hspace{-1.9mm} /} - M)\psi
\end{eqnarray}
after integrating the fermions we obtain (\ref{p140}). Observe that there is no need to deal with singular gauge transformations so that the fermionic determinant is well defined. As far as we know this computation of the fermionic determinant subject to singular potentials is a new result that corroborates the Julia-Toulouse condensation computation.

In conclusion we have discussed a new approach to deal with the MCS theory in the presence of magnetic defects based on the Julia-Toulouse condensation mechanism. This mechanism, which can be seen as dual to the perturbative radiative corrections in this instance, leading naturally to the appearance of the CS term, is here shown to be able to generate also the {\it ad hoc} HT electric current present in the magnetic brane, as a natural consequence of the condensation process and overall consistence of the formulation. As a result, the magnetic brane becomes {\it observable}. This is the explicit signature of confinement of the instantons bordering the brane. As an immediate consequence electric confinement is destroyed. Furthermore the {\it physical reality} of the brane may possibly interfere with the delicate Aharonov-Bohm process that quantizes the CS coefficient. In fact, our results suggest that this system might contain anyonic degrees of freedom. This is a new possibility that might have an impact on effective planar systems in condensed matter such as the multilayer Hall fluid or Josephson junctions. In fact, recently we were able to apply the Julia-Toulouse rationale into the field-localization problem \cite{Dvali:2007nm}, where Josephson junction ideas are used to localize gauge fields in a brane. Using the methodology developed in this paper we were able to consider also the inclusion of matter (fermionic) fields leading to the destruction of the IR long rang confinement found in earlier works \cite{localization}.

 \section{Acknowledgments}

The authors would like to thank Funda\c{c}\~ao
de Amparo \`a Pesquisa do Estado do Rio de Janeiro (FAPERJ) and Conselho Nacional de
Desenvolvimento Cient\' ifico e Tecnol\'ogico (CNPq) and CAPES (Brazilian agencies)
for financial support.

 \end{document}